\documentclass[12pt]{article}
\usepackage{amssymb,amsmath}
\begin{document}

\begin{titlepage}

                            \begin{center}
                            \vspace*{5mm}
\Large\bf{Escort   \     distributions   \     and   \   Tsallis   \   entropy}\\

                            \vspace{2.5cm}

              \normalsize\sf    NIKOS \  \  KALOGEROPOULOS $^\S$\\

                            \vspace{2mm}
                            
 \normalsize\sf Weill Cornell Medical College in Qatar\\
 Education City,  P.O.  Box 24144\\
 Doha, Qatar\\

                            \end{center}

                            \vspace{2.5cm}

                     \centerline{\normalsize\bf Abstract}
                     
                           \vspace{3mm}
                     
\normalsize\rm\setlength{\baselineskip}{18pt} 

\noindent  We present an argument justifying the origin  of the escort distributions used in 
calculations involving the Tsallis entropy. We rely on an induced hyperbolic Riemannian metric reflecting 
the generalized composition property of the Tsallis entropy.
The mapping of the corresponding Riemannian measure  on the space of thermodynamic variables gives 
the specific form of the escort distributions and 
provides a geometric interpretation of the non-extensive parameter. In addition, we explain the 
polynomial rate of increase of the sample space volume for systems described 
by the Tsallis entropy, thus extending the previously reached conclusions for discrete systems
to the case of systems whose evolution is described by flows on  Riemannian manifolds.\\

                             \vfill

                             
                             \vfill

\noindent\rule{8cm}{0.2mm}\\
   \noindent \small\rm $^\S$  E-mail: \ \  \small\rm nik2011@qatar-med.cornell.edu\\

\end{titlepage}


                                                                                 \newpage

 \normalsize\rm\setlength{\baselineskip}{18pt}

The Tsallis entropy [1], [2], in the case of a system described by a continuous sample space \ $\Omega$ \ endowed with a 
probability distribution \ $\rho (x), \ x\in\Omega$, \ is given by  
\begin{equation}
S_q [\rho]  = k_B \frac{1}{q-1} \left\{1 - \int_{\Omega} [\rho(x)]^q \ dvol_{\Omega} \right\}
\end{equation}
where \ $q\in\mathbb{R}$ \ is called entropic, or non-extensive, parameter. The  Boltzmann/Gibbs/Shannon (BGS) entropy 
\begin{equation}
        S_{BGS} [\rho] = - k_B \int_{\Omega} \rho(x) \log \rho(x) \ dvol_{\Omega} 
\end{equation}
is recovered for \ $q\rightarrow 1$. \ We set henceforth the Boltzmann constant \ $k_B = 1$, \ for simplicity. For two 
independent, as are conventionally defined, subsystems \ $\Omega_1,  \Omega_2 \subset \Omega$ \ by definition 
\begin{equation}  
      \rho_{\Omega_1 \ast \Omega_2} = \rho_{\Omega_1} \rho_{\Omega_2}
\end{equation}
where \ $\Omega_1 \ast \Omega_2$ \ indicates the combined system arising from the interactions of \ $\Omega_1$ \ and  
\ $\Omega_2$. \ We  immediately see that the Tsallis entropy (1) is not additive for this definition of independent systems, but it instead obeys
\begin{equation}
      S_q (\Omega_1 \ast \Omega_2) = S_q (\Omega_1) + S_q (\Omega_2) + (1-q) S_q(\Omega_1) S_q(\Omega_2) 
\end{equation}
To make the Tsallis entropy explicitly additive, one [3], [4] re-defines the concept of ``independence" by introducing what is essentially a 
modified Abelian group structure whose addition, reflecting (4), is given by 
\begin{equation}
    x \oplus_q y = x + y + (1-q)xy  
\end{equation}
and whose multiplication [5], [6]  is given by
\begin{equation}
x \otimes_q y = \frac{1}{1-q} \cdot  \left\{(2-q)^\frac{ \log[1+(1-q)x]  \log[1+(1-q)y]}{[\log (2-q)]^2} - 1\right\} 
\end{equation} 
These two operations (5), (6) introduce a field structure on a deformation of \ $\mathbb{R}$, \ indicated by \ $\mathbb{R}_q$,  \ 
which is explicitly given by the field isomorphism [6]
\begin{equation}
    \tau_q (x) =  \frac{(2-q)^x - 1}{1-q} 
\end{equation}
 
For reasons of invertibility discussed in [6], we will henceforth confine our attention to the 
subset of non-extensive parameters \ $q\in [0, 1] \subset \mathbb{R}$. \ 
In order to get a better idea of the meaning and properties of the generalized definition of additivity (5),  in [7] we compared
side by side \ $\mathbb{R}$ \ and \ $\mathbb{R}_q$. \  This comparison was implemented by setting up and contrasting appropriately defined 
metrics  induced by the ordinary and generalized (5) additions of \ $\mathbb{R}$ \ and \ $\mathbb{R}_q$ \  respectively. 
We started by using the usual flat Euclidean metric on the plane \ $\mathbb{R}^2$,  \  expressed in Cartesian \ $(x, y)$ \ coordinates by the metric 
tensor 
\begin{equation}   
{\bf g} = \left(    \begin{array}{ll} 
                                 1 & 0 \\
                                 0 & 1\\ 
                                \end{array}         \right)
\end{equation}
Then, by using a semi-direct product construction based on (5), hence on the Tsallis entropy composition property (4), 
we found a metric \ ${\bf g}_h$ [6] \ that encodes the comparison of the ordinary and the generalized additions. Expressed in the 
same \ $(x,y)$ \ Cartesian coordinate system, \ ${\bf g}_h$ \ has components 
\begin{equation}
   {\bf g}_h = \left(    \begin{array}{ll} 
                                 1 & 0 \\
                                 0 & e^{-2tx}\\ 
                                \end{array}         \right)
\end{equation}
It was seen in [7] that (9) has constant negative sectional curvature
\begin{equation} 
 k = - \{ \log (2-q) \}^2
\end{equation}
where the parameter \ $t$ \ in (9) is related to  the nonextensive parameter \ $q$ \ by 
\begin{equation}
    t = \log (2-q) 
\end{equation}
The above constructions can be immediately extended to the case of \ $\mathbb{R}^n$, \ instead of just \ $\mathbb{R}^2$, \  generalizing (9) to 
\begin{equation} 
      {\bf g}_h = \mathrm{diag} (1, \underbrace{e^{-2tx_1}, \ldots, e^{-2tx_1}}_{n-1} )
\end{equation}
where \ $x = (x_1, \ldots, x_n) \in \mathbb{R}^n$. \ It is obvious that to zeroth order (8) and (9) are equal, as they should be, since both are 
Riemannian metrics hence both Euclidean (``flat") to zeroth order. Curvature is a second order effect, after all, as the first-order deviations from 
flatness appear to be encoded in the connection coefficients (Christoffel symbols). The latter, being non-tensorial objects, hence devoid of any 
coordinate-independent interpretation, are zero in appropriately chosen coordinates. Alternatively, a simple counting argument shows that setting
the Christoffel symbols to zero is akin to ``gauge fixing" of the metric. As a second order approximation though, the deviation from flatness 
encoded in the Riemann curvature tensor acquires a geometric, coordinate-free, meaning.
 Let's indicate \ $\mathbb{R}^n$ \ endowed with (12) by \ $\mathbb{H}^n (t)$ \ 
with \ $t$ \ given by (11),  and the  invertible map between (8) and (9) by \ $h: \mathbb{R}^n \rightarrow \mathbb{H}^n(t)$ 
\begin{equation}
h({\bf g}) = {\bf g}_h 
\end{equation}
 
When we usually describe a dynamical system with a locally Euclidean space as sample space \ $\Omega$, \  we tacitly assume that 
the composition property (3) holds. This composition reflects the conventional definition of independence. Hence a metric analysis of \ $\Omega$ \  
starts by using the extension of (8) to \ $\mathbb{R}^n$. \ If it turns out that the system's collective behavior is described by the Tsallis entropy, 
and moreover if this is somehow reflected through the generalized composition property (5)  at the level of \ $\Omega$, \ 
then the system should actually be described by the effective metric (12) as was established in [7].
Hence it is (12) rather than the initially and straightforwardly, but naively, chosen extension of (8) which should be used in the 
analysis of the system under study. Based on this realization, and using the above construction, we showed in a previous work  [8] that 
the largest Lyapunov exponent of the underlying dynamical system described by the Tsallis entropy should vanish a statement which is  in 
agreement, so far as we know today, with all evidence arising from the analysis of specific models [2].\\

We presently  continue alongside this line of development,  by exploring some measure-theoretical, rather than metric as was done in [8], 
consequences of the hyperbolic nature of the effective metric (12) induced by the composition property of the Tsallis entropy (4).  
To make matters a bit more concrete,  we define a map 
\begin{equation}
    j: \mathbb{R}^n \rightarrow T_x\Omega
\end{equation}
where the sample space \ $\Omega$ \ is assumed to be a Riemannian manifold, \ $x\in\Omega$, \ with its tangent space at \ $x$ \ 
being indicated by \ $T_x\Omega$ \ and \ $T\Omega = \cup_{x\in\Omega} T_x\Omega$ \ indicating the tangent bundle of \ $\Omega$. \   
The evolution of the dynamical system takes place in \ $\Omega$, \ $T\Omega$ \ or yet another vector bundle with base \ $\Omega$ \ depending on 
the system. The main examples of such spaces are the configuration and phase spaces \ $M$ \ of  many-body  systems. 
Here \ $n\in \mathbb{N}$ \ can be treated as an, initially, free parameter. A simple choice is to  consider \ $j$ \ to be the identity map between 
\ $\mathbb{R}^n$ \ and \ $T_x\Omega$ \ which would amount to  \ $n = \dim \Omega$. \
Then one obtains the metric \ ${\bf g}_\Omega$ \ through the derivative of the exponential map \ $d\exp : \mathbb{R}^n \rightarrow T_x\Omega$. \ 
More generally, \ $n$ \ can be chosen to be sufficiently large and \ $j$ \ to be continuous so that
\begin{equation}
   J  ({\bf g}) \equiv (d\exp \circ j ) \  {\bf g} = {\bf g}_\Omega    
\end{equation}
By a similar procedure, the effective metric (12) can be used to define the $j$-``hyperbolized" map, in analogy with (14), by 
\begin{equation}
  j_h: \mathbb{H}^n (t) \rightarrow T_x\Omega
\end{equation}
and the hyperbolic analogue of (15) by
\begin{equation}
 J_h ({\bf g}) \equiv (d\exp \circ j_h \circ h) \  {\bf g}
\end{equation}
giving
\begin{equation} 
  J_h ({\bf g}) = ({\bf g}_\Omega)_h
\end{equation} 
In  words, what \ $J_h$ \ does, is to take \ $T_x\Omega$ \ and substitute in each of these tangent spaces the metric \ ${\bf g}$ \ 
of \ $\mathbb{R}^n$ \  with that of  \ $\mathbb{H}^n(t)$ \ (12). \  One way to construct \ $({\bf g}_\Omega)_h$ \ is by using a partition of unity 
and following the solvable group  construction that lead from (8) to (9), for each coordinate patch. 
As an indication of the first steps of such a construction, 
we start by expressing the Euclidean metric  of \ $\mathbb{R}^n$ \ in polar coordinates as
\begin{equation}   
     ds^2 = dr^2 + r^{n-1} \ du^2
\end{equation}
where \ $r$ \ stands for the distance along the radial geodesics and \ $du^2$ \ indicates the usual line element  of the 
Euclidean \ $(n-1)$-dimensional sphere \ $\mathbb{S}^{n-1}$ \ of unit radius. This can be re-written, in a warped product notation, as 
\begin{equation}
   ds^2 = dr^2 \times_{r^{n-1}} \ du^2
\end{equation}
Using geodesic normal coordinates on \ $\Omega$ \ with radial parameter \ $r$, \ one can locally express the line element of \ ${\bf g}_\Omega$ \ as
\begin{equation}
  ds^2 = dr^2 \times_{r^{n-1}} \ dv^2
\end{equation}
where \ $dv^2$ \ is the line element on the geodesic sphere of radius \ $r$ \ in \ $\Omega$.  \ In complete analogy, and with similar notation, to the 
(19), (20), by using the Poincare ball model for \ $\mathbb{H}^n(1)$, \ one has 
\begin{equation} 
   ds^2 = dr^2 \times_{\sinh^{n-1}r} \ du^2 
\end{equation}
which upon mapping on \ $\Omega$ \ gives the metric \ $({\bf g}_\Omega)_h$ \ in polar coordinates as  
\begin{equation}  
    ds^2 = dr^2 \times_{\sinh^{n-1}r} \ dv^2 
\end{equation}
As stated above this is just an indication of how to construct such metrics locally. However, gluing the local coordinate patches together on 
\ $\Omega$ \ in a consistent way so as to  define globally \ $({\bf g}_\Omega)_h$ \ is a non-trivial, and not even always possible, step. We will 
tacitly assume from now on that such a (model-dependent) construction is possible.
Except for the next paragraph that addresses a potential topological obstruction, we will not elaborate on its details
as they are not needed for the rest of our argument. \\

One set of possible obstructions to the above construction may come from topology. Consider as a typical example the restrictions that the  
Gauss-Bonnet theorem places on properties of admissible metrics on Riemannian surfaces. As an immediate application, one sees that a torus does 
not admit a smooth everywhere negative or everywhere positive sectional curvature metric. 
In our case, we notice that our arguments pertain to the behavior of the Riemannian volume \ $dvol_{\Omega}$ \ and of the measure \ $d\mu$ \ which 
is absolutely continuous with respect to \ $dvol_{\Omega}$, \ which  arises from the ``hyperbolization" of \ ${\bf g}_\Omega$. \  
The behavior  \ $dvol_{\Omega}$ \ is certainly controlled by the sectional curvature of \ ${\bf g}_{\Omega}$, \ but imposing conditions 
on the sectional curvature of \ $({\bf g}_\Omega)_h$ \ is unnecessarily restrictive. What is pertinent for our purposes is  an understanding of the 
properties of the Ricci curvature of \ ${\bf g}_\Omega$ \ which  controls the behavior of \ $dvol_{\Omega}$. \  The Ricci curvature determines this 
behavior of \ $dvol_{\Omega}$ \  through a Riccati equation [9], [10]. In Physics, this approach has been of considerable significance  in proving  
singularity theorems of General Relativity. This happens because  the trace of that Ricatti equation is the Raychaudhuri equation which is 
instrumental  (in Lorentzian signature spaces though) in proving the alluded to singularity theorems [11].   So, it is not necessary to focus in finding 
topological obstructions constructing a negative sectional curvature metric \ $({\bf g}_\Omega)_h$ \ on \ $\Omega$ \ 
but rather at obstructions to constructing negative Ricci curvature metrics. To this end, however,  a theorem of Lohkamp [12] states that a
Riemannian manifold of dimension at least  three,  always admits a metric of negative Ricci curvature irrespective of its topology. Hence, no 
topological obstructions exist in the construction of a suitable class of metrics. Hence,  so it is indeed possible to construct  an appropriate
for our purposes \ $({\bf g}_\Omega)_h$, \ at least in principle.\\

Since the evolution of the dynamical system on \ $\Omega$ \ is described by the effective metric (18), the corresponding 
effective measure of interest  is not the Riemannian volume \ $dvol_{\Omega}$ \ but   
\begin{equation}
      d\mu = J_h  \left(   \mathrm{Jac}  \ dvol_{\Omega}   \right)
\end{equation}
where \ $\mathrm{Jac}$ \ stands for the Jacobian of the map \ $J_h  \circ J^{-1}$ \ according to the area formula [13].  
Using the Poincare ball model of \ $\mathbb{H}^n(1)$ \ alongside (15) and (17) we see that 
\begin{equation}
      \mathrm{Jac} = e^{-\{(n-1)t \} r}  
\end{equation}
giving 
\begin{equation}
   d\mu = J_h  \left(  e^{-\{(n-1)t \} r}   \ dvol_{\Omega}   \right)
\end{equation}
In order to proceed we need some additional information about (18) or, equivalently, about (17). A very simple choice for 
\ $J_h$ \ would be to be such that it maintains the form of the measure, possibly only changing the value of \ $n$. \ An example of such a map 
is provided by a subset of monomial maps having exponent a rational number \ $w\in \mathbb{Q}$. \ 
This is a convenient choice, as such maps have simple scaling properties but still have a rich enough structure that allows them to describe 
many important phenomena associated to self-similar structures [14],  as the latter provided a motivating set of examples for the initial 
introduction and development of the  Tsallis entropy [1], [2].  Examples of such maps, in low dimensions, are the complex quadratic map,
and if lower degree polynomial forms are also allowed then the category includes  Arnold's cat map, the tent map, the logistic map etc [15]. 
Such maps, for \ $r>0$ \  have a leading-power monomial behavior given by
\begin{equation}  
       r \ \mapsto \ r^w, \hspace{8mm} w\in\mathbb{Q}
\end{equation}
When (27) is combined with (26) for such maps, it gives for \ $n' \in \mathbb{N}$
\begin{equation}
     d\mu = e^{-\{ (n'-1)t \} r} J_h( dvol_{\Omega})
\end{equation}

The transition from the microscopic evolution of the dynamical system's configuration/phase space \ $M$ \ to its thermodynamic description amounts, 
in the simplest case, to defining  maps
\begin{equation}
    p: (M, \ ({\bf g}_M)) \rightarrow (\tilde{M}, {\bf \tilde{g}})
\end{equation}
and its hyperbolic counterpart     
\begin{equation}
    p_h: (M, \ ({\bf g}_M)_h)) \rightarrow (\tilde{M}, {\bf \tilde{g}}_h) 
\end{equation}
where \ $\tilde{M}$ \ is the space of thermodynamic variables, quite often also assumed to be a Riemannian manifold,
 \ ${\bf \tilde{g}}$ \ is its ``Euclidean"  metric and \ ${\bf \tilde{g}}_h$ \ its the hyperbolic analogue of \ ${\bf \tilde{g}}$. \ 
There have been several constructions of such ``thermodynamic" metrics \ ${\bf \tilde{g}}$ \ pertaining to systems described 
by the BGS entropy over the years. Two such notable cases are the Weinhold [16] and Ruppeiner [17], [18] metrics, for instance. 
One of the simplest ways that the non-trivial composition property of the Tsallis entropy  (4) can be reflected in the thermodynamics,
in this formalism, is by requiring the effective pushforward measure \ $(p_h)_{\ast} d\mu$ \ on \
$\tilde{M}$ \ to have the same relation to the Riemannian measure of \ $\tilde{M}$, \ as the relation of \ $d\mu$ \ with \ $dvol_M$. \ 
This reflects that the self-averaging of the microscopic system resulting from its statistical description, that lends validity to its thermodynamic,
behavior is rather mild. This  means that, as an assumption, the self-averaging should not give rise to substantially new emergent features.
Although it may not be so obvious how to make this statement more precise, at this stage, an example of the opposite behavior may be illuminating.
Such a contrast can be drawn, for instance, with the case of spin glasses [19]. 
In spin glasses the co-existence of disorder and frustration gives rise to non-trivial geometric structures. If one works in the mean field 
approximation of the Kirkpatrick-Sherrington model, for instance, and uses the replica method with the Parisi ansatz, then one discovers that 
there is a very rich landscape of local energy minima [19]. This landscape has a hierarchical tree structure within this approximation, and is best 
described by ultrametrics [20], in sharp contrast to the Euclidean inner product of the spin degrees of freedom employed in the definition of the 
Kirkpatrick-Sherrington Hamiltonian.   Going back to our argument, under this assumption about the lack of  radical emergent behavior upon 
self-averaging, and with \ $dvol_{\tilde{M}}$ \ indicating the volume element of \ $\tilde{M}$ \ with respect to \ ${\bf \tilde{g}}$, \   the effective 
pushforward measure on \ $\tilde{M}$ \ induced by (4), is by an extension of (28), given by  
\begin{equation}
    (p_h)_{\ast} d\mu = e^{-(N-1) t R} \ dvol_{\tilde{M}}
\end{equation}
Here \ $\dim (p_h)_{\ast}\mu = N$ \ stands for the pointwise dimension of \ $(p_h)_{\ast} d\mu$, \ defined as [15], [21] 
\begin{equation} 
     \dim \ (p_h)_{\ast} \mu (\tilde{x}) \ = \ \lim_{R\rightarrow 0} \ \frac{\log \ \ (p_h)_{\ast} \mu(B(\tilde{x}, R))}{\log \ R}
\end{equation}
where \ $B(\tilde{x}, R)$ \ stands for the ball with center \ $\tilde{x}$ \ and of radius \ $R>0$, \ in \ $(\tilde{M}, {\bf \tilde{g}}_h)$. \ 
The naive expectation is that the Hausdorff measure \ $dvol_{\tilde{M}}$ \ in the $N$-dimensional space \ $\tilde{M}$ \ should transform 
multiplicatively, by  a factor of \  $e^{-NtR}$, \ under  a conformal change of metric. However \ ${\bf \tilde{g}}_h$ \ is not derived from 
\ ${\bf \tilde{g}}$ \ from such a conformal change of the metric, but rather from the analogue of the ``hyperbolizaton" map (18) on \ $\tilde{M}$, \                            
which reflects the difference in the definitions of  ``independence" and consequently of addition induced by the BGS and the Tsallis entropies.  
As a result, \ ${\bf \tilde{g}}_h$ \ actually varies according to (31), namely it  
acquires  a multiplicative factor of  \ $e^{-(N-1)tR}$. \ If \ $N\in\mathbb{N}$, \ (31) would  be interpreted  as stating that \ $(p_h)_{\ast} d\mu$ \ 
corresponds to an area ($(N-1)$-dimensional Hausdorff  measure) in \ $\tilde{M}$ \ rather than a volume ($N$-dimensional Hausdorff measure). 
For \ $N\not\in\mathbb{N}$ \  the interpretation is extended by analogy, now however referring to the pointwise dimension of the measure 
\ $(p_h)_{\ast} d\mu$ \ instead of the Hausdorff dimension(s) of various (sub-)sets in \ $\tilde{M}$. \ We would like to repeat that this 
discrepancy between the naively expected and the actual dimensions in the conformal factor in (31), reflects the difference between 
the ordinary and generalized additions induced by the BGS and the Tsallis entropies, respectively. 
Alternatively, (31) can be interpreted as  quantifying at the thermodynamic level,  the different definitions of ``independence" employed by the BGS 
and the Tsallis entropies in describing the collective behavior of the systems of  interest. \\

We now briefly digress to present a related comment about  the ergodicity of systems described by the Tsallis entropy.
A corollary of Birkhoff's ergodic theorem [15], states that if the system's evolution can be represented by an ergodic, measure-preserving  flow 
\ $f_t: M \rightarrow M$, \ with respect to measure \ $\nu$, \ 
then for any  integrable function  \ $\phi \in L^1(\Omega)$ \  
\begin{equation}
      \lim_{T\rightarrow\infty} \ \int_0^T \phi(f_t(x)) \  dt = \int_{\Omega} \phi(x) \ d\nu 
\end{equation}
This holds true for almost all \ $x\in M$, \ namely all \ $x$ \ outside a set of \ $\nu$ \ measure zero on \ $M$. \ In particular, this result can be applied 
to the case of a Hamiltonian system by using as \ $\nu$ \ the Liouville measure which, if the system has simple enough topology and interactions, 
can be reduced to the Riemannian (Sasakian) volume element \ $dvol_M$ \ on the space of its generalized coordinates \ $M$. \  
In addition, such an ergodic  probability measure $\nu$, \ whose Radon-Nikodym derivative with respect to the Lebesgue measure of \ $M$ \ will be 
denoted by  \ $\rho(x)$, \ is unique. To be more precise, if \ $\rho_1$ \ and \ $\rho_2$ \ are two such ergodic, invariant (probability) densities, 
then they should be singular with  respect to each other.\\
 
Equilibrium statistical mechanics largely relies on the existence of such a stationary ergodic measure,  stemming from  Birkhoff's 
ergodic theorem, which provides the equilibrium ensemble \ $\rho(x)$ \ with respect to which all macroscopic quantities of interest are 
calculated. For any such  quantity \ $\mathcal{A}$, \ one can calculate its average as
 \begin{equation}
     \langle \mathcal{A} \rangle  = \int_M \ \mathcal{A}(x) \rho(x) \ dvol_M
\end{equation}
 In the case of non-equilibrium systems the Sinai-Ruelle-Bowen measures seem to play a similar role, but the extent of their genericity 
 and scope of applicability is still somewhat unclear. The maximization of the BGS entropy subject to the appropriate thermodynamic constraints gives 
 rise to the desired form of \ $\rho(x)$ \ for each of the classical equilibrium ensembles (microcanonical, canonical etc). By contrast [2], it is claimed that 
 the Tsallis entropy is useful in  cases when the system's evolution is not ergodic. This statement tacitly assumes that ``independence" is defined in the 
 conventional way (3), or equivalently, that evolution  of the system on \ $M$ \ is described by (15) and its induced Riemannian measure 
 \ $d\nu = dvol_M$ \ on \ $M$. \  As it becomes obvious by a comparison between the Shannon [22] and Santos [23] axioms, or alternatively, between 
 the Khinchin [24] and Abe [25] axioms,  the difference between the BGS and Tsallis entropies amounts to  a difference between the ordinary and the 
 generalized addition (5). These different definitions of addition quantify the two different ways of thinking about ``independence" for systems 
 described by the BGS and the Tsallis entropies as noticed above. From a metric viewpoint, this is initially reflected in the difference 
 between (8) and (9) and the induced map (13). Eventually this boils down to the difference between the thermodynamic description of the system by 
\ ${\bf \tilde{g}}$ \ and \ ${\bf \tilde{g}}_h$ via (29) and (30) respectively. \ Alluding to each of the two pairs of axioms mentioned above, it becomes 
clear that the evolution of a  system described by the Tsallis entropy becomes ergodic, hence one can still apply the Birkhoff ergodic theorem in 
equilibrium cases, but now with respect to \ $d\mu$ \ (28) instead of the originally, naively but arbitrarily, chosen Riemannian volume \ $dvol_M$. \\

 At the level of thermodynamics, the stationary distribution \ with respect to \ ${\bf \tilde{g}}$ \ has density given by \ $\rho \sim L^{-N}$ \ where \ $L$ \ 
 represents a small length on \ $\tilde{M}$, \ but it is not related to an ergodic evolution of the system, as noted above. On the other hand, 
 with respect to the effective metric \ ${\bf \tilde{g}}_h$, \ we see from (31) and the arguments in the previous paragraph, that the corresponding 
 density \ $\rho_h$ \ is ergodic and behaves as  \ $\rho_h \sim L^{-(N-1)}$. \ Therefore, we have found that
\begin{equation}      
      \rho_h = \rho^\frac{N-1}{N}
\end{equation} 
is the density with respect to which the evolution is of the system ergodic. It is therefore \ $\rho_h$ \ in (35) that should be used in 
calculating the values of any quantities of statistical interest of the underlying dynamical system, rather than the initially expected \ $\rho$. \   Set 
\begin{equation} 
   q = \frac{N-1}{N}
\end{equation} 
indicating by \ $q$ \ the entropic/non-extensive parameter appearing in the definition (1) of the Tsallis entropy. Since \ $N\in [1, +\infty)$, \ we see that \ 
$q\in [0,1]$. \  What we have accomplished is to have justified the use of the escort distribution \ $\rho_h = \rho  ^q$, \ (35) instead of the naively 
expected \ $\rho$, \ in the calculations of macroscopic quantities of systems described by the Tsallis entropy. 
We have also ascribed the origin of the non-extensive parameter \ $q$ \  (36) to the pointwise dimension  \ $N$ \ of the effective measure 
\ $(p_h)_\ast \ d\mu$ \ (31). \\ 
 
An immediate consequence of the above construction is an alternative justification of the well-known fact that the Tsallis entropy describes systems
whose configuration/phase space volume increases in a polynomial/power-law manner [26], [27], [2]
\begin{equation}
    Vol_\Omega \sim n^s,  \hspace{5mm} s\in\mathbb{N}
\end{equation}
with the number of degrees of freedom \ $n$ \ of the system. We can contrast this growth rate with the exponential manner
\begin{equation}
   Vol_\Omega \sim a^n,  \hspace{5mm}  a\in (1, +\infty)
\end{equation}
in which the corresponding volume grows for systems described by the BGS entropy. 
Indeed, if the system were described the the BGS entropy, then its behavior would be metrically captured by the 
Euclidean metric ${\bf g}_\Omega$ (15). In that case its configuration/phase space would increase exponentially with the 
number of degrees of freedom according to (38). Let us, however, assume that this system is actually described by the Tsallis, 
instead of the BGS, entropy. Then the corresponding stationary probability distribution  will have ``fat tails" which give rise to 
long-range spatial and temporal correlations for the system. As a result, the accessible part of the 
configuration/phase space is severely restricted [26], [27], [2].  Such a behavior  is now quantified by the effective 
hyperbolic metric \ ${\bf \tilde{g}}_h$ \ (18) and its associated infinitesimal volume element  (28), 
which varies with the number of effective degrees of freedom as  
\begin{equation} 
     \mu \sim e^{n''-1}   
\end{equation} 
with \ $n'' \sim n$ \ due to (26), (28). Therefore, and for \ $n, n'' \gg 1$, \ we have  
\begin{equation}
    \mu \sim Vol_{\Omega} 
\end{equation}
So from the viewpoint of the effective measure \ $\mu$, \ the actual volume \ $Vol_\Omega$ \ can increase at most sub-exponentially as 
both have to have the same leading exponential growth rate in terms of the number of the effective degrees of freedom (40).  
This condition is trivially accomplished if each volume varies in a polynomial/power-law way with respect to \ $n$, \ 
which is exactly the sought after result.  This conclusion was previously reached in the case of (binary) discrete systems [26], [27], [2]. 
Our argument above extends the conclusion to the 
case of  systems  described by the Tsallis entropy having evolution modelled by a flow on a Riemannian manifold. It also justifies why the escort, 
\ $\rho_h$ \ rather than the naively expected distributions \ $rho$ \ should to be used in formulating the thermodynamic constraints, 
in the derivation of the stationary distributions, when applying the maximum entropy approach for the Tsallis entropy [28], [29]. \\

A question of possible interest  is to determine a tensorial quantity on \ $(\Omega, {\bf g}_\Omega)$ \ that controls locally the variations 
of \ $d\mu$, \ in a similar manner as  the Ricci tensor of \ ${\bf g}_\Omega$ \ controls the variations of \ $dvol_\Omega$, \ via a Riccati equation 
as noted above. The Ricci tensor is used extensively in geometric descriptions of Statistical Mechanics [30] associated to systems described by the 
BGS entropy. A corresponding quantity for statistical systems described by the Tsallis entropy should generalize the Ricci tensor and since it would 
be a local object, it would have the advantage of being relatively simple to explicitly calculate in the models of interest.\\
 
Another question that arises naturally from the above arguments, is to examine whether there are systems whose configuration/phase 
space volume growth rate with respect to the number of degrees of freedom is faster than polynomial/power-law, yet slower than exponential. 
The argument culminating in (40) allows for the possibility of sub-exponential but super-polynomial rates of growth of the volume of \ $M$. \ 
Systems with such behavior would have an intermediate growth rate in their configuration/phase space volumes, placed between (37) and (38). 
Do such systems exist, and if so, are they described by the BGS, the Tsallis or yet some other more ``intermediate" form of entropy? 
The naive expectation based on  \ $\lim_{q\rightarrow 1}  S_q = S_{BGS}$ \ is that such ``intermediate" behavior should not exist because 
the Tsallis entropy describes all the possible non-BGS behavior if one works within the framework of the axioms [23], [25]. 
On the other hand, given the vast array of systems that the  Tsallis entropy purports to describe [2], many of which are not modelled on Riemannian 
spaces, may allow for some possible surprises in this direction. To achieve this without contradicting the axioms [23], [25],  someone might try to 
relax somewhat or generalize the composition property (4) and check what is possible, and probably search among the nonextensive families of 
entropies determined by two or more nonextensive parameters [2] which include the Tsallis entropy as a special case. 
This question is clearly motivated, but seems not to be related at first glance, to a formally similar 
question [31]  about the growth rate of finitely generated groups endowed with the word metric. An answer to the question posed in [31] was   provided 
by the Grigorchuk  group [32], [33] whose  construction spurred considerable fruitful activity in geometric group theory. In a similar 
manner, a potentially non-trivial answer to the analogous question formulated above for the case of entropies and the 
configuration/phase space volume growth rates may be the starting point for a better understanding of the dynamical basis of the BGS, the 
Tsallis and of possibly other entropies. \\

                                                                          \vspace{10mm}

\noindent {\large\sc Acknowledgement:} \ We wish to thank Professor Constantino Tsallis for pointing out the similarities and differences of 
maximizing the Tsallis entropy under thermodynamic constraints calculated with respect to the original $\rho$ and to the escort distributions 
$\rho_h$,  and for bringing to our attention references [28] and [29] where these matters are extensively discussed.\\

                                                                           \vspace{12mm}

                                                                                                                         
                                                        \centerline{\large\sc References}
 
                                                                          \vspace{5mm}
 
\noindent [1]  C. Tsallis, \  \emph{J. Stat. Phys.} {\bf 52}, \ 479 \ (1988)\\
\noindent [2]  C. Tsallis, \  \emph{Introduction to Nonextensive Statistical Mechanics: Approaching a Complex\\
                            \hspace*{4mm} World},  \ Springer (2009)\\
\noindent [3] L. Nivanen, A. Le Mehaut\'{e}. Q.A. Wang, \ \emph{Rep. Math. Phys.} {\bf 52}, \ 437 \ (2003)\\
\noindent [4] E.P. Borges, \ \emph{Physica A} {\bf 340}, \ 95  \ (2004)\\
\noindent [5] T.C. Petit Lob\~{a}o, P.G.S. Cardoso, S.T.R. Pinho, E.P. Borges, \ \ \emph{Braz. J. Phys.} {\bf 39}, \\
                             \hspace*{4mm}  402 \ (2009)\\
\noindent [6] N. Kalogeropoulos, \ \emph{Physica A} {\bf 391}, \ 1120 \ (2012)\\
\noindent [7] N. Kalogeropoulos, \ \emph{Physica A} {\bf 391}, \ 3435 \ (2012)\\
\noindent [8]  N. Kalogeropoulos, \emph{Vanishing largest Lyapunov exponent and Tsallis entropy}, {\sf arXiv:1203.2707}\\ 
\noindent [9] H. Karcher, \ \emph{Riemannian Comparison Constructions}, \ Global Diff. Geom., \ Vol. 27, \\ 
                             \hspace*{4mm} Math. Assoc. Amer. (1989)\\
\noindent [10] J.-H. Eschenburg, E. Heintze, \ \emph{Manuscr. Math.} {\bf 68}, \ 209 \ (1990)\\
\noindent [11] S.W. Hawking, G.F.R. Ellis, \ \emph{The large scale structure of space-time}, \ Cambridge \\
                             \hspace*{6mm}  University Press \ (1973)\\ 
\noindent [12]  J. Lohkamp, \  \emph{J. Diff. Geom.} {\bf 40}, \ 461 \  (1994)\\ 
\noindent [13] H. Federer, \  \emph{Geometric Measure Theory}, \ Springer (1969)\\
\noindent [14] A.F. Beardon, \ \emph{Iteration of Rational Functions}, \ Springer (1991)\\
\noindent [15] C. Beck, F. Schl\"{o}gl, \ \emph{Thermodynamics of chaotic systems: an introduction}, \ Cambridge \\
                            \hspace*{6mm} University Press (1993)\\
\noindent [16] F. Weinhold, \ \emph{J. Chem. Phys.} {\bf 63}, \ 2479 \ (1975)\\
\noindent [17] G. Ruppeiner, \ \emph{Phys. Rev. A} {\bf 20}, \ 1608 \ (1979)\\
\noindent [18] G. Ruppeiner, \ \emph{Rev. Mod. Phys.} {\bf 67}, \ 605 \ (1995)\\ 
\noindent [19] M. Mezard, G. Parisi, M.A. Virasoro, \ \emph{Spin Glass Theory and Beyond}, \ World Scientific \\
                            \hspace*{6mm}  (1987)\\
\noindent [20] R. Rammal, G. Toulouse, M.A. Virasoro, \ \emph{Rev. Mod. Phys.} {\bf 58}, \ 765 \ (1986)\\
\noindent [21] Y.B. Pesin, \ \emph{Dimension Theory in Dynamical systems: Contemporary Views and \\
                             \hspace*{6mm}  Applications}, \ University of  Chicago Press (1997)\\ 
\noindent [22] C. Shannon, \ Bell. Syst, Tech. Jour. {\bf 27}, \ 379 (1948); \ \  Ibid. {\bf 47}, \  623 (1948)  \\
\noindent [23] R.J.V. Santos, \ \emph{J. Math. Phys.} {\bf 38}, \ 4104 \ (1997)\\ 
\noindent [24] A.I. Khinchin, \ \emph{Mathematical Foundations of Information Theory}, \ Dover (1957) \\ 
\noindent [25] S. Abe, \  \emph{Phys. Lett. A} {\bf 271}, \ 74 \ (2000)\\
\noindent [26] C. Tsallis, M. Gell-Mann, Y. Sato, \ \emph{Proc. Natl. Acad. Sci.} {\bf 102}, \ 15377 \ (2005)\\
\noindent [27] R. Hanel, S. Thurner, \ \emph{Europhys. Lett.} {\bf 96}, \ 50003 \ (2011)\\ 
\noindent [28] W.J. Thisleton, J.A. Marsh, \ \emph{IEEE Trans. Infor. Th.} {\bf 53},  \ 4805 (2007)\\ 
\noindent [29] C. Tsallis, A.R. Plastino, R.F. Alvarez-Estrada, \ \emph{J. Math. Phys.} {\bf 50}, \ 043303 \ (2009)\\
\noindent [30] L. Casetti, M. Pettini, E.G.D. Cohen, \ \emph{Phys. Rep.} {\bf 337}, \ 237 \  (2000)\\
\noindent [31] J.W. Milnor, \ \emph{J. Diff. Geom} {\bf 2}, 447 \ (1968)\\
\noindent [32] R.I. Grigorchuk, \ \emph{Funct. Anal. Appl.} {\bf 14}, \ 41 \ (1980)\\
\noindent [33] R.I. Grigorchuk, \ \emph{Sov. Math. Dokl.} {\bf 28}, \ 23 \ (1983)
\end{document}